\documentclass[twocolumn,superscriptaddress,aps,prb,amsmath,preprintnumbers]{revtex4-2}
\usepackage{tikz}
\usepackage{graphicx}
\usepackage{hyperref}
\usepackage{epsfig}
\usepackage{xcolor}
\usepackage{amsmath, amssymb}
\usepackage[left,pagewise]{lineno}

\definecolor{darkgreen}{rgb}{0,0.7,0}
\newcommand{\TK}[1]{\textcolor{black}{#1}}

\usepackage{epstopdf}

\begin{document}

%\linenumbers
%\pagewiselinenumbers
%-------------------------------------------------------------------
%\title{Transient crystal.... in a linear chain}
%\title{Charge waves and time dynamics in SSH topological chains}  
%\title{Charge waves and transient dynamics as probes of the topological phase in SSH chains}
%\title{Charge waves and transient dynamics \TK{distinguishing} topological phases in SSH chains}
\title{Charge waves and dynamical signatures of topological phases in Su–Schrieffer–Heeger chains}

\author{T.\ Kwapi\'nski}
\affiliation{Institute of Physics, M.\ Curie-Sk\l odowska University, 20-031 Lublin, Poland }
\email{tomasz.kwapinski@umcs.pl} 
\author{M. Kurzyna}
\affiliation{Department of Computer Science and Mathematics, M. Curie-Sklodowska University, 20-031  Lublin, Poland}
\author{L. E. F. Foa Torres}
\affiliation{Departamento de F\'{\i}sica, Facultad de Ciencias F\'{\i}sicas y Matem\'aticas, Universidad de Chile, Santiago, Chile}

%\address{%
%	$^{1}$ \quad Department of Physics, Maria Curie-Sklodowska University in Lublin, Poland; \\
%	$^{2}$ \quad Department of Comp. Sci., Maria Curie-Sklodowska University in Lublin, Poland;

      \date{\today}
%------------------------------------------------------------------

\begin{abstract}
We investigate the emergence of charge waves and their temporal dynamics in one-dimensional Su–Schrieffer–Heeger (SSH) topological chains. Contrary to the conventional view that charge oscillations are suppressed in gapped topological systems with preserved chiral symmetry, we show that such oscillations can indeed occur. 
\TK{The general condition for an arbitrary oscillation period is analysed, and we find that the charge waves propagating along the chain do not depend on its topology, except at the edges, where both topological phases exhibit essential differences. In chains with inequivalent atoms within the SSH unit cell, we observe regular long-period sublattice oscillations that appear simultaneously with even–odd charge oscillations.} 
Furthermore, we study the nonequilibrium dynamics in SSH chains. After a quench, the time evolution of the local density of states and charge occupancies exhibits clear dynamical fingerprints that distinguish topologically trivial and nontrivial phases. 
\TK{Our results establish that transient charge dynamics can distinguish topologically trivial and nontrivial phases in real time by detecting the presence of topologically-protected edge states.}
\end{abstract}

%It is believed that our results  will bring new perspectives to a wide range of time crystal physics. They can be verified experimentally using  time spectroscopy techniques or in photonic crystals.


%\pacs{73.23.-b,73.21.La}

% 73.23.-b  Electronic transport in mesoscopic systems
% 73.21.La  Quantum dots
% 72.15.Qm  Scattering mechanisms and Kondo effect
% 74.45.+c  Proximity effects; Andreev reflection; SN and SNS junctions

%  \textcolor{red}{stepped}

\maketitle

%%%%%%%%%%%%%%%%%%%%%%%%%%%%%%%%%%%%%%%%%%%%%%%%%%%%%%%%%%%%%%%%%%%%%%
\section{\label{sec1}Introduction}
%%%%%%%%%%%%%%%%%%%%%%%%%%%%%%%%%%%%%%%%%%%%%%%%%%%%%%%%%%%%%%%%%%%%%%%

% PARAGRAPH 1 (Original - Opening)
Advances in nanotechnology have enabled the study of one-dimensional (1D) systems such as linear quantum dots, quantum wires, and atomic chains coupled to electron reservoirs. These structures represent the ultimate limit of miniaturization in electronic transport and offer promising applications in spintronics, quantum computing, and precision metrology \cite{Crain2004, Baski2001,jal}.
Due to the confinement of electron motion, such atomic-scale systems exhibit a range of unique physical phenomena, including spin-charge separation, conductance oscillations, charge waves or fractional charges \cite{Spin01,Kwap2011,x39,Yeom2022,Yeom2022NN,kwap2005} to name just a few. 

% PARAGRAPH 2 (Original - Context)
In particular, spatial charge oscillations in low-dimensional structures, which arise from the wave nature of electrons, are the main subject of this work. In 1D systems, these oscillations are commonly referred to as Friedel oscillations \cite{Fri1952,Fri1958,Dobson1995}, which result from electron scattering due to impurities or boundary confinement \cite{Byczuk2015,Dora2005,Maska2007}. These oscillations reflect the underlying electronic structure and can reveal details of the band structure, including Berry phases, as demonstrated in multilayer graphene systems \cite{Dutr2016}.

% PARAGRAPH 3 (REVISED - Model Definition + Experimental Platforms)
Recently, topological materials have emerged as a prominent research area due to their ability to support boundary states within an energy gap while acting as insulators in the bulk. One of the most elementary realizations is the Su-Schrieffer-Heeger (SSH) model \cite{SSHx01,SSHx02,SSHx03,SSHx11}, originally proposed to describe polyacetylene. The model features a dimerized chain with alternating intra- and inter-dimer couplings, supporting two distinct topological phases: nontrivial (SSH1) with edge states localized at both chain ends, and topologically trivial (SSH0) without edge states but with an energy gap throughout. 
Nontrivial topology can be also observed in extended SSH models featuring various geometries,
such as long-range chains incorporating next-nearestneighbor hoppings or by multi-site unit cells \cite{SSHx11,x002,Torres2024,SSHx03}. 
SSH-type systems have been realized in various experimental platforms, including quantum optics \cite{x002,x007,x008}, vacancy defects in chlorine superstructures on Cu(100) \cite{Drost2017, Huda2020}, indium-decorated silicon atomic chains on Si(553)-Au substrates \cite{Jal2024}, and arrays of quantum dots \cite{Pham2022}.

% PARAGRAPH 4 (Original - The Gap / Challenge)
Despite extensive studies of SSH chains, two fundamental questions remain unresolved. First: \textit{Can charge  waves emerge in gapped SSH topological systems with preserved chiral symmetry?} 
Conventional understanding suggests that such oscillations arise from a modulation of the local density of states at the Fermi level along the chain and therefore should be suppressed in materials with an energy gap around $E_F$. Note that the emergence of charge density waves breaks the translational symmetry of the system, which can typically occur due to the breaking of chiral symmetry by a local instability or electron–phonon coupling.
Second, while static topological properties are well-characterized through winding numbers and edge state spectroscopy, the \textit{dynamical signatures} that distinguish topological phases during nonequilibrium evolution have received considerably less attention. Dynamical probes could provide experimentally accessible alternatives to traditional spectroscopic methods, particularly relevant for atomic-scale systems where time-resolved measurements may be more feasible than direct band structure mapping.

% PARAGRAPH 5 (Original - The Resolution / "In this work...")
In this work, we address both questions. 
Regarding the first question, we demonstrate that charge oscillations \textit{can} occur in gapped topological systems through two distinct mechanisms: (i) when the bulk bands cross the Fermi energy, even with preserved chiral symmetry, and (ii) through explicit breaking of chiral symmetry via inequivalent sublattices. We show that the oscillation periods are governed by the average site occupancy and that these waves can be observed in both topologically trivial and nontrivial phases, though with distinct characteristics at the chain edges.
Regarding the second question, we uncover \TK{a dynamical signature that distinguishes topological phases:} following a sudden quench of coupling parameters, the topologically nontrivial phase exhibits transient oscillations with two distinct timescales that directly reflect the emergence of topological edge states within the bulk gap. In contrast, the trivial phase shows uniform oscillations with identical periods at all atoms throughout the chain. 
\TK{These distinct oscillation timescales provide an experimentally accessible signature of the topologically protected edge states whose existence is guaranteed by the nontrivial winding number ($\nu = 1$). This offers a real-time method for distinguishing topological phases that complements conventional spectroscopy. We further show that transient charge dynamics provide a sensitive means of detecting transitions between topological phases.}

% PARAGRAPH 6 (Former P7 - Broader Context for Dynamics)
The dynamical aspects of this work connect to studies of time dynamics in driven low-dimensional conductors, where exotic nonequilibrium phases have emerged, such as time crystals \cite{x03,x04,x05} and transient crystals \cite{kwap2020}. Previous work has explored a rich spectrum of quantum effects, including photon-assisted tunneling \cite{Wiel2002,Kwap2004,x14}, spin and charge quantum pumping \cite{05,22,Kwap2011}, transient current beats \cite{Souza2007, x12}, and Floquet topological phases \cite{x01,x02,Lago}. However, the specific transient signatures that can signal different topological phases following a quench remain less explored.

% PARAGRAPH 7 (Former P8 - Organization)
The paper is organized as follows. In Sec.~\ref{sec2}, we describe the theoretical model and the Hamiltonian, \TK{ while the technical details of the calculation methods for both the stationary and time-dependent cases are presented in Appendix~\ref{app:calc}. Section~\ref{sec4} discusses the main results of the paper, focusing on charge waves with different oscillation periods in SSH chains and on the time evolution of the local density of states and charge occupancies in the system. Finally, the last section provides a brief summary of our findings. The role of the system geometry is discussed in Appendix~\ref{app:geometry}. }

%%%%%%%%%%%%%%%%%%%%%%%%%%%%%%%%%%%%%%%%%%%%%%%%%%%%%%%%%%%%%%%%%%%%%%
\section{\label{sec2} Theoretical description}
%%%%%

To analyze the charge distribution along the chain we consider a system consisting of $N$ atoms arranged in a linear configuration on a substrate or coupled to individual electrodes, which act as electron reservoirs. Within the tight-binding framework, the system is described by the Hamiltonian $H=H_0+H_{coupl}$, where  $H_0$ represents the single-electron part, 
\TK{ and $H_{coupl}$ accounts for coupling Hamiltonian. In general, Hamiltonian may be time dependent; however, electron–electron  interactions are not included in our model.} Explicitly:
%
%%%%%%%%%%%%%%%%%%%%%%%%%%%%%%%%%%%%%%%%%%%%%%%%%%%%%%%%%%%%%%%%%%%%%%
\begin{eqnarray} 	\label{001} 
	H_{0}&=& \sum_{i=1}^N \varepsilon_i c^{\dagger}_i c_i + \sum_{\vec{k}}  \varepsilon_{\vec{k}} c^{\dagger}_{\vec{k}} c_{\vec{k}} \,, \\
	H_{coupl} &=& \sum_{i=1}^N\sum_{\vec{k}} V_{\vec{k} i} c^{\dagger}_{\vec{k}} c_{i} + \sum_{i=1}^{N-1} t_{i,i+1}(t) c^{\dagger}_i c_{i+1} + H.c. \nonumber\,, 
\end{eqnarray}
%%%%%%%%%%%%%%%%%%%%%%%%%%%%%%%%%%%%%%%%%%%%%%%%%%%%%%%%%%%%%%%%%%%%%%
%
where $\vec{k} = \vec{k}_{\text{leads}}$ represents the electron wave vectors of the leads (electrodes), which are coupled to the atoms in the chain via hybridization elements $V_{\vec{k} i}$. We consider two basic geometries: one in which each atom of the chain is coupled to the surface electrode, and the L-R geometry, where the electron reservoirs are connected only to the end atoms of the chain. In the latter case, the inner atoms are located on an insulating or semiconducting substrate. Such a system can be realized either using atoms placed on vicinal surfaces or by fabricating a linear series of quantum dots with fully controlled coupling parameters. 
The operators $c^{\dagger}_{i/\vec{k}}$ and $c_{i/\vec{k}}$ denote the electron creation and annihilation operators for $i$ or $\vec{k}$ quantum state.
The nearest-neighbor chain sites are coupled via the hopping integrals $t_{i,i+1}(t)$, which in general are time-dependent. 

%\textbf{Stationary Hamiltonian.}

In the stationary case, the system Hamiltonian is time-independent, and the couplings between atoms take constant values.
The SSH topological chain is characterized by two alternating couplings: between sites within a two-atom unit cell, and between neighboring cells, denoted as $t_{i,i+1} = t_1$ (for odd $i$) and $t_{i,i+1} = t_2$ (for even $i$), respectively, \cite{SSHx01,SSHx02,SSHx03}, and in general one can write $t_{i,i+1} = (t_1, t_2)$. The chain with $t_1 = t_2$ corresponds to a regular, uniform chain without topological states and with no energy gap. For $t_1 < t_2$, the system enters a nontrivial topological phase (SSH1), exhibiting an energy gap and mid-gap states localized at both ends of the chain. Conversely, for $t_1 > t_2$, the chain is in a topologically trivial phase (SSH0), with no edge topological states. The SSH chain is characterized by the topological winding number $\nu$, which takes values $\nu = 0$ for $t_1 > t_2$ (trivial phase, SSH0) and $\nu = 1$ for $t_1 < t_2$ (nontrivial phase, SSH1)~\cite{SSHx01}. When $\nu = 1$, the bulk-boundary correspondence guarantees the existence of a topological edge state at energy $E = \epsilon_i$ localized at both chain ends. 
Both phases exhibit an energy gap of width $2|t_1-t_2|$, and two bulk bands (sidebands) each with a width of $2 \min(t_1,t_2)$. 
The key distinction between the two phases lies not only in the presence or absence of edge states, but also in how the system responds to perturbations, as we will demonstrate in the following sections through analysis of charge distributions and dynamical evolution.

\TK{Electronic properties of a stationary system is analyzed within the framework of Green's functions \cite{Datta,Des}, for details see Appendix~\ref{app:calc}.
For the time-dependent Hamiltonian, the interatomic couplings vary in time, $t_{i,i+1}(t)$, which induces a temporal evolution of the LDOS at each atomic site. To describe these dynamical processes, we employ the interaction picture and the evolution-operator formalism \cite{33,29,kwap2020}, as detailed also in Appendix A.
}

In our calculations, we use $\Gamma_0 \equiv 1$ as the unit of energy, with time measured in units of $\hbar/\Gamma_0$. For $\Gamma_0 = 1$~meV, this corresponds to a time unit of approximately $0.66$~ps. The reference energy is set at the surface Fermi level, $E_F = 0$, and all calculations are performed at zero temperature. 
Throughout this work, we adopt the following notation: time is denoted by $t$, the initial time by $t_0$ (with $t_0=0$), and the interatomic couplings $t_{i,i+1}$ are denoted by $t_i$ for a normal chain or by $(t_1,t_2)$ for topological SSH chains.

%\newpage
%%%%%%%%%%%%%%%%%%%%%%%%%%%%%%%%%%%%%%%%%%%%%%%%%%%%%%%%%%%%%%%%%%%%
\section{\label{sec4}Results and discussions}
%%%%%%%%%%%%%%%%%%%%%%%%%%%%%%%%%%%%%%%%%%%%%%%%%%%%%%%%%%%%%%%%%%%%

%%%%%%%%%%%%%%%%%%%%%%%%%%%%%%%%%%%%%%%%%%%%%%%%%%%%%%%%%%%%%%%%%%%%
\subsection{\label{sec4a} Charge waves in SSH chains}
%%%%%%%%%%%%%%%%%%%%%%%%%%%%%%%%%%%%%%%%%%%%%%%%%%%%%%%%%%%%%%%%%%%%

In regular atomic chains, the electronic density of states spans the entire energy band within the range of $\pm 2 t_i$, exhibiting spatial site-to-site modulations that facilitate the formation of charge waves along the system \cite{Kwap2006, Jaloch2023}. In contrast, topological chains feature a fundamentally different DOS, characterized by an energy gap around the Fermi level and the presence of topologically protected edge modes. This raises a fundamental question: can charge waves still emerge in such a system? \TK{Before presenting our results, we clarify terminology. The 'charge waves' or 'charge oscillations' discussed here are Friedel-type oscillations arising from quantum interference at system boundaries, analogous to those observed near impurities and surfaces in metals. These should not be confused with Peierls-type charge density waves (CDW), which involve electron-phonon coupling and spontaneous symmetry breaking. Our charge oscillations are properties of the noninteracting electronic system and do not require symmetry breaking or lattice instabilities.}
%
%%%%%%%%%%%%%%%%%%%%%%%%%%%%%%%%%%%%%%%%%%%%%%%%%%%%%%%%%%%%%%%%%%%%%%
\begin{figure}[tb]
	\begin{center}
		\includegraphics[angle=0,width=0.9\columnwidth]{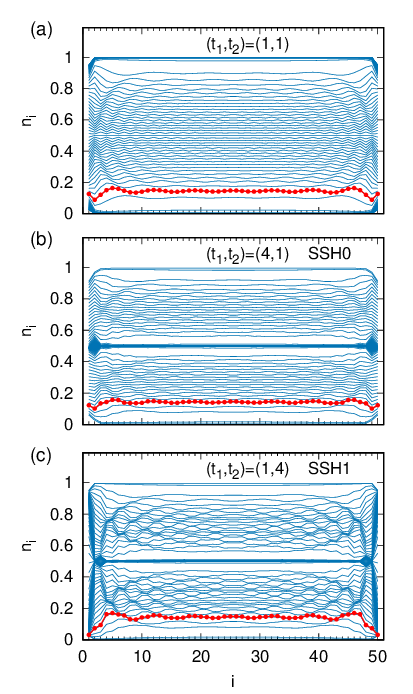}
	\end{center}
	\caption{Charge occupancies at each site $i$ for a chain of length $N=50$, shown for three cases: a normal chain ($t_i=1$, panel a),  SSH chain in the trivial phase ($t_1=4$, $t_2=1$, SSH0, panel b), and  SSH chain in the nontrivial phase ($t_1=1$, $t_2=4$, SSH1, panel c). The curves in each panel correspond to different on-site energies, ranging from $\varepsilon_i=-5$ (upper curves) to $\varepsilon_i=+5$ (bottom curves) with a step of $0.2$. The system is in the L-R geometry, meaning $\Gamma_i=0$ everywhere except at the boundaries, where $\Gamma_1=\Gamma_N=1$. \TK{The red curves  correspond to the oscillation period $M=7$ i.e. $\varepsilon_i=+1.80$ (upper panel), and $\varepsilon_i=+4.69$ (middle and bottom panels). }
		 The reference energy point is the lead Fermi energy, $E_F=0$, energy is expressed in $\Gamma_0$ units.  } \label{charge01}
\end{figure}
%%%%%%%%%%%%%%%%%%%%%%%%%%%%%%%%%%%%%%%%%%%%%%%%%%%%%%%%%%%%%%%%%%%%%%
%
%Fig.~\ref{charge01} ...
To address this question, Fig.~\ref{charge01} presents the charge distributions along various chains (normal, SSH0 and SSH1 chains) for different values of the on-site energy $\varepsilon_i$. 
In each panel, the lower curves correspond to $\varepsilon_i = +5$, while the upper curves represent $\varepsilon_i = -5$, and the intermediate curves are plotted for values of $\varepsilon_i$ between these extremes, with a step of 0.2. 
\TK{This figure is intended to illustrate the overall highly regular pattern formed by the family of such curves for the SSH chain (middle and lower panels) and to show its relation to the pattern obtained for the regular chain (upper panel).}
The top panel illustrates the case of a normal atomic chain, in 
which all hopping integrals  along the chain are equal. We observe that the charge curves tend to grouping/cluster at specific chain sites, forming denser (darker) regions due to overlapping curves. This clustering leads to 
oscillations in charge occupancy along the chain for certain values of $\varepsilon_i$. As is well known, these oscillations appear under the condition defined by relation:  
 $\cos\left( \frac{\pi l}{M}\right)=\frac{E_F-\varepsilon_{i}}{2t_i}$, where $M$ is the oscillation period and $l=1,...,M-1$ \cite{kwap2005,Kwap2006}. Thus e.g. the oscillations of period $M=3$, 4 or 5 atoms emerges for $\varepsilon_i=\pm t_i$,  
$\varepsilon_i=\pm \sqrt{2} t_i$, or  $\varepsilon_i=\pm t_i(1\pm \sqrt{5})/2$, respectively. Notably, in the presence of charge  waves in a normal chain, the oscillation period is directly related to the average charge occupancy in the system: for a period $M$, the average charge occupancy is $<n>=1/M$ \cite{Kwap2006}. 

The situation is different for a topological SSH chain, as shown in panel (b) for the trivial phase and panel (c) for the nontrivial phase. In the trivial SSH0 chain, an energy gap of width $2|t_2-t_1|=6$ spans the entire system. 
For such values of  $\varepsilon_i$ that the Fermi energy lies within the gap, the chiral symmetry of the system enforces a uniform site occupancy of $n_i=0.5$, resulting in a dense clustering of curves at these  $\varepsilon_i$. For other values of $\varepsilon_i$, however, one of the LDOS sidebands crosses the Fermi energy, leading to a modulation of the LDOS along the chain. As a result, well-defined charge-occupancy patterns appear for both positive and negative $\varepsilon_i$ (see also  Fig.~\ref{chiral1} and Fig.~\ref{geometry} for a detailed analysis of the local DOS distributions in the SSH chain). These oscillations give rise to charge waves in the topological system, analogous to those in a uniform chain.

A comparable behavior of charge waves is observed for the SSH1 chain in the nontrivial phase (bottom panel, c). Here, two distinct regions of charge oscillations emerge, associated with the LDOS sidebands, which resemble the charge distribution in a normal chain (panel a) but are twice as narrow. Additionally, in the nontrivial phase the grouping of the charge curves is slightly more pronounced and sharper than in the trivial chain shown in panel (b); \TK{ however, the main differences appear at both ends of the chain due to the presence of topological edge states. Furthermore, owing to the topologically protected end modes,} the charge occupancy at the end sites $i=1$ and $i=N$ is relatively low for $\varepsilon_i>0$ and high for $\varepsilon_i<0$, which manifests as densely shaded regions at both edges of the chain. This effect will play a crucial role in the analysis of the time dynamics in chains with different topological phases. \TK{The red curves in all panels of Fig.~\ref{charge01} represent an example of the charge-occupation profile along the chain for the oscillation period $M=7$; these will be discussed later. }

It is also important to highlight the role of symmetry in the studied systems. For all systems considered in Fig.~\ref{charge01}, the charge distribution along the chain is always spatially symmetric, i.e.  $n_{i}=n_{N-i+1}$, and antisymmetric with respect to the position of the single-particle energy level $\varepsilon_i$ relative to the Fermi energy $E_F=0$. In other words, each charge occupancy profile $n_i$ corresponding to a given $\varepsilon_i$ is related to the profile for $-\varepsilon_i$ by a reflection about the reference value $n_i=0.5$, which constitutes a direct manifestation of the chiral symmetry.

%\newpage
%\begin{widetext}
%
%%%%%%%%%%%%%%%%%%%%%%%%%%%%%%%%%%%%%%%%%%%%%%%%%%%%%%%%%%%%%%%%%%%%%%
\begin{figure}[tb] 
	\begin{center}
		\includegraphics[angle=0,width=0.85\columnwidth]{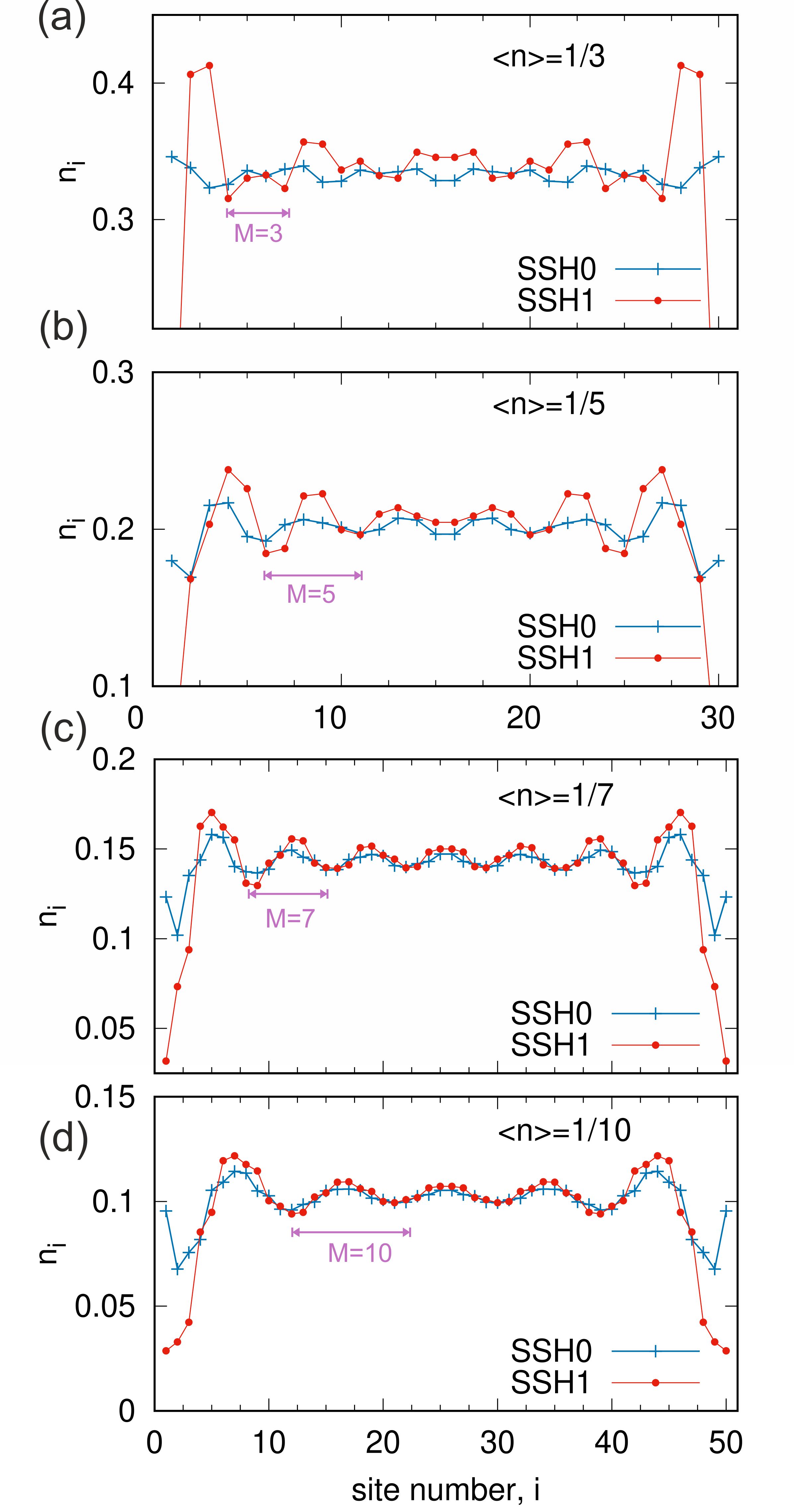}
	\end{center}
	\caption{Charge occupancies along a chain of length $N=30$ (panels a and b) or $N=50$ (panels c and d) sites in the L-R geometry, for values of the on-site energies $\varepsilon_i$ (the same for all chain sites) corresponding to oscillation periods of $M=3$ (panel a, $\varepsilon_i=3.6$), $M=5$ (panel b, $\varepsilon_i=4.41$), $M=7$ (panel c, $\varepsilon_i=4.69$) and $M=10$ (panel d, $\varepsilon_i=4.84$), as indicated by arrows in the plots. The oscillation periods are related to the average charge values in the chain and for panels (a–d) are $<n>=1/3, 1/5, 1/7, 1/10$, respectively.  Blue and red lines represent the chain in the trivial ($t_1=4$, $t_2=1$), and in the nontrivial ($t_1=1$, $t_2=4$) phase. }
	\label{charge02}
\end{figure}
%%%%%%%%%%%%%%%%%%%%%%%%%%%%%%%%%%%%%%%%%%%%%%%%%%%%%%%%%%%%%%%%%%%%%%
%\end{widetext}
% Fig.~\ref{charge02}.
In contrast to a normal chain, in topological SSH systems, we observe a slightly different, split/bifurcated charge distribution structure. We will analyze in more detail the formation of such waves in both topologically trivial and nontrivial chains. It turns out that for a chain in the SSH geometry, the condition for the existence of charge waves with a period of $M$ sites can be written as:
\begin{eqnarray} \label{condition_SSH}
 E_F - \varepsilon_{i}=\pm \sqrt{t^2_1+t^2_2 +2t_1t_2 \cos\left(\frac{2\pi l}{M}\right) } \,.
\end{eqnarray} 
Since the LDOS width of each SSH sideband is $2\min(t_1,t_2)$ and these sideands are shifted by $\pm \max(t_1,t_2)$ relative to the Fermi level, the values of $\varepsilon_i$ corresponding to an oscillation period of $M$ sites are located such that the Fermi level lies within the sideband region of the chain DOS. 
\TK{Note that the system topology is not directly relevant here, as the charge oscillations are related to the sideband structure of the spectrum, which is independent of the topological phase. Only the enhancement of the oscillation amplitude at the edges in the SSH1 phase reflects the presence of edge states.}

In Fig.~\ref{charge02}, we analyze charge waves with periods of $M = 3$, 5, 7, and 10, corresponding to panels (a), (b), (c), and (d), respectively. The red lines represent the chain in the topologically nontrivial phase, while the blue lines correspond to the trivial phase. Charge oscillations along the chain arise when the LDOS exhibits spatial modulations at the Fermi level. In the present case, this condition is satisfied by tuning the on-site energies $\varepsilon_i$ such that the lower LDOS sideband crosses the Fermi level. The specific values of $\varepsilon_i$ used in panels (a)–(d) are obtained from Eq.~\ref{condition_SSH} and are 3.6, 4.41, 4.69 \TK{(see also the red curves in  Fig.~\ref{charge01})}, and 4.84, respectively. 
It is worth noting that the same values of $\varepsilon_i$ are obtained when they are determined self-consistently by requiring that the average site occupancy along the chain be fixed by the charge-oscillation period $M$, i.e., $\langle n \rangle = 1/M$. Such self-consistent calculations of $\varepsilon_i$ ensure that this constraint is satisfied. The numerically obtained values of $\varepsilon_i$ for large $N$ are in excellent agreement with those predicted by the analytical relation.
% between the average occupancy and the charge-wave period.

In each case shown in Fig.~\ref{charge02}, the oscillation amplitude at the chain ends is  larger than in the middle of the chain, which is a characteristic feature of Friedel oscillations. This effect is more pronounced in the nontrivial phase (SSH1), where the end-chain oscillation amplitude is stronger, as also seen in Fig.~\ref{charge01}. 
These differences arise because, in the SSH1 chain, the topological states draw spectral weight from the sidebands, resulting in slightly reduced LDOS values in the sidebands compared to the SSH0 chain. Consequently, for positive $\varepsilon_i$   the occupation of the end sites in the SSH1 phase is significantly suppressed relative to the trivial phase. Thus, the presence of charge-wave modulations with large amplitudes at the chain ends \TK{reflects the spectral weight redistribution due to edge states, and is characteristic of the nontrivial phase. In the trivial phase, where no edge states exist, oscillation amplitudes at the boundaries are more moderate.} For shorter chains this effect may become even more pronounced; however, in such systems well-developed charge waves do not form.

\TK{We have also verified that similar effects and charge waves in both SSH0 and SSH1 topological chains can be observed in a linear geometry, where all sites are coupled to the surface, similarly to what is observed in the left–right (L–R) geometry, where only the end sites of the chain are coupled to electron reservoirs; for details, see Appendix~\ref{app:geometry}.}

%%%%%%%%%%%%%%%%%%%%%%%%%%%%%%%%%%%%%%%%%%%%%%%%%%%%%%%%%%%%%%%%%%%%

%%%%%%%%%%%%%%%%%%%%%%%%%%%%%%%%%%%%%%%%%%%%%%%%%%%%%%%%%%%%%%%%%%%%
\subsection{\label{sec4c} Broken chiral symmetry}
%%%%%%%%%%%%%%%%%%%%%%%%%%%%%%%%%%%%%%%%%%%%%%%%%%%%%%%%%%%%%%%%%%%%

An isolated SSH atomic chain with identical on-site energies $\varepsilon_{i}$ possesses chiral symmetry regardless of the absolute value of $\varepsilon_{i}$ and a rigid shift of these levels does not break this symmetry. However, in the situation considered here, the reference point is the Fermi energy of the electron reservoir (substrate or L/R electrodes). Consequently, a shift of the energy levels leads to the breaking of the symmetry of the LDOS function  on each atom with respect to  $E_F$, which, as we discussed in Fig.~\ref{charge01} and Fig.~\ref{charge02}, may give rise to charge oscillations along the chain.

%
%%%%%%%%%%%%%%%%%%%%%%%%%%%%%%%%%%%%%%%%%%%%%%%%%%%%%%%%%%%%%%%%%%%%%%
\begin{figure}[tb]
	\begin{center}
		\includegraphics[angle=0,width=0.99\columnwidth]{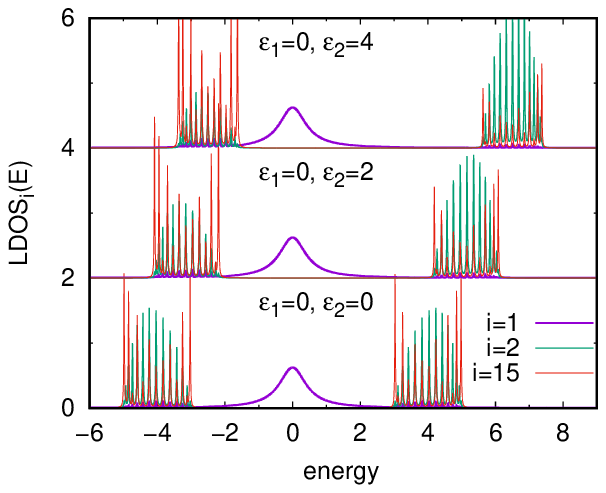}
	\end{center}
	\caption{LDOS at three sites $i=1,2,15$ (violet, green, and red lines, respectively) of the SSH1 chain ($t_1=1$, $t_2=4$) with length $N=30$. The on-site energies in the unit cell are: $\varepsilon_1=0$, and $\varepsilon_2=0,2,4$ (bottom, middle, and upper curves, respectively), i.e. in the chain $\varepsilon_i=\varepsilon_1$ for odd $i$, and $\varepsilon_i=\varepsilon_2$ for even $i$. For clarity, the middle and upper curves are shifted upward by $+2$ and $+4$. }
	\label{chiral1}
\end{figure}
%%%%%%%%%%%%%%%%%%%%%%%%%%%%%%%%%%%%%%%%%%%%%%%%%%%%%%%%%%%%%%%%%%%%%%
%\end{widetext}
One mechanism for inducing LDOS asymmetry in the chain, and thus breaking chiral symmetry, is to consider an SSH chain with a two-site cell composed of different atoms. In this case, the symmetry is broken due to different energy values of the states in the cell, $\varepsilon_{1}$ and $\varepsilon_{2}$, 
which define two inequivalent sublattices with $\varepsilon_i=\varepsilon_1$ for odd $i$, and $\varepsilon_i=\varepsilon_2$ for even $i$. Figure~\ref{chiral1} presents an example of the LDOS for  three atoms in an SSH1 chain consisting of $N=30$ sites. The lower curves correspond to the case of equal on-site energies, $\varepsilon_{1}=\varepsilon_{2}$, for which the LDOS is fully symmetric with respect to the Fermi level. When the on-site energies are different ($\varepsilon_{1}=0$ and $\varepsilon_{2}=2,4$, upper curves), the LDOS sidebands shift toward higher energies, while the topological state remains pinned at $\varepsilon_{1}=0$.
The LDOS function then becomes strongly asymmetric. For instance, atom $i=2$ (green curves) exhibits enhanced spectral weight above the Fermi energy, whereas the odd atom $i=15$ (red curves) exhibits stronger LDOS below $E_F$. As a result, upon calculating the charge distribution along the chain (by integrating the LDOS up to $E_F$), sites with lower single-particle energies acquire larger occupancies compared to those with higher energies $\varepsilon_i$. This effect gives rise to a charge wave with period $M=2$, i.e., even–odd charge oscillations. These oscillations originate from chiral-symmetry breaking due to the presence of two inequivalent atoms in the unit cell of the SSH1 chain (which also occurs in an analogous manner in the SSH0 chain).

\begin{figure}[tb]
	\begin{center}
		\includegraphics[angle=0,width=0.9\columnwidth]{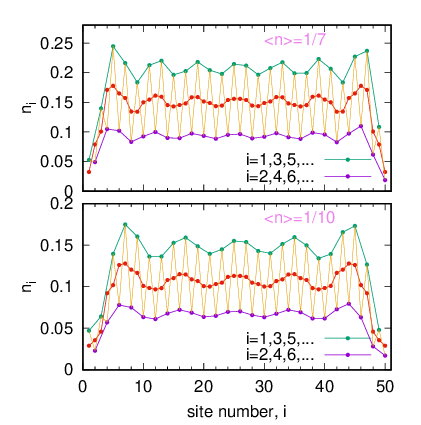}
	\end{center}
	\caption{Charge occupancies along the SSH1 chain of length $N=50$ in the L–R geometry ($t_1=1$, $t_2=4$). The red curves correspond to equal on-site energies $\varepsilon_1=\varepsilon_2$ and exhibit oscillation periods $M=7$ (upper panel) and $M=10$ (lower panel), identical to those in Fig.~\ref{charge02}. The yellow curves correspond to broken chiral symmetry with $\varepsilon_2-\varepsilon_1=4$, i.e., $\varepsilon_{1,3,5,\ldots}=3.07$ and $\varepsilon_{2,4,6,\ldots}=7.07$ (upper panel), and $\varepsilon_{1,3,5,\ldots}=3.22$ and $\varepsilon_{2,4,6,\ldots}=7.22$ (lower panel). The green and violet lines show charge oscillations in each sublattice separately. The oscillation periods are determined by the average filling of the chain, with $\langle n\rangle=1/7$ and $\langle n\rangle=1/10$, respectively.}
	\label{chiral2}
\end{figure}

It is therefore natural to ask whether charge density waves with periods larger than 2 can emerge in such a system, still within a two-site cell composed of different atoms. To obtain charge oscillations with an arbitrary period, it is necessary for a LDOS sideband to cross the Fermi level.  For the SSH1 chain with broken chiral symmetry we performed self-consistent calculations of the charge distribution for periods $M=7$ (upper panel in Fig.~\ref{chiral2}) and $M=10$ (lower panel in Fig.~\ref{chiral2}).
%
%%%%%%%%%%%%%%%%%%%%%%%%%%%%%%%%%%%%%%%%%%%%%%%%%%%%%%%%%%%%%%%%%%%%%%
%%%%%%%%%%%%%%%%%%%%%%%%%%%%%%%%%%%%%%%%%%%%%%%%%%%%%%%%%%%%%%%%%%%%%%
%\end{widetext}
%
The red lines correspond to the symmetric case $\varepsilon_{1}=\varepsilon_{2}$ (as in Fig.~\ref{charge02}, panels c and d), where oscillations with $M=7$ and $M=10$ are clearly observed. In contrast, the yellow lines in Fig.~\ref{chiral2} correspond to the case with broken chiral symmetry ($\varepsilon_{2}-\varepsilon_{1}=4$), for which  the $M=2$ even–odd charge oscillations are well visible. Remarkably, when the charge densities of the two sublattices are plotted separately (i.e., considering every second atom in the chain), both sublattices display oscillations with the same spatial periods as in the symmetric chain (see the green and violet envelopes).

Thus, in the two-sublattice scenario, chiral symmetry is broken from the outset (two different atoms in the unit cell), and consequently charge oscillations with period $M=2$ necessarily emerge. 
However, when the energy levels are further shifted so that an LDOS sideband crosses the Fermi level (thus breaking the energy symmetry with respect to $E_F$) additional sublattice oscillations with longer periods also appear. Note that such sublattice oscillations have not been reported in the literature before.

%%%%%%%%%%%%%%%%%%%%%%%%%%%%%%%%%%%%%%%%%%%%%%%%%%%%%%%%%%%%%%%%%%%%
\subsection{\label{sec4d} Time dynamics of charge waves}
%%%%%%%%%%%%%%%%%%%%%%%%%%%%%%%%%%%%%%%%%%%%%%%%%%%%%%%%%%%%%%%%%%%%

In this chapter, we analyze the time evolution of charge occupations in an atomic chain following a sudden change in the coupling  parameters. Initially, the chain is in a normal (equilibrium) state with uniform coupling between neighboring sites. Subsequently, the stationary state is perturbed by a sudden change in the site-to-site couplings, transforming the system into a topological SSH chain characterized by alternating coupling strengths $(t_1, t_2)$.
During this process, the system remains in a nonequilibrium state for a finite time, while the chain’s DOS undergoes a dynamic reconstruction. Eventually, the system relaxes into the topological phase.

%
%%%%%%%%%%%%%%%%%%%%%%%%%%%%%%%%%%%%%%%%%%%%%%%%%%%%%%%%%%%%%%%%%%%%%%
\begin{figure}[tb]
	\begin{center}
		\includegraphics[angle=0,width=0.85\columnwidth]{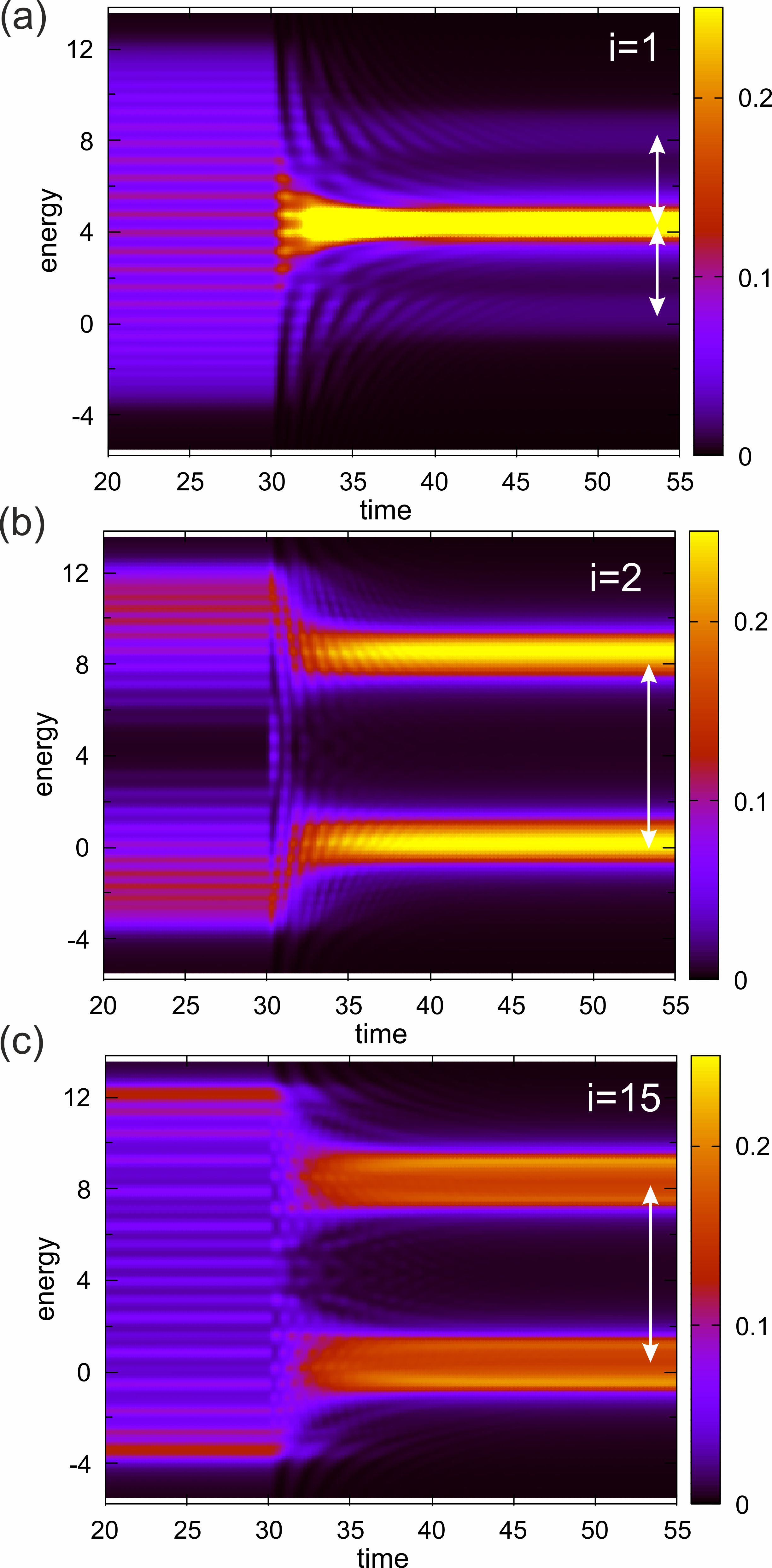}
	\end{center}
	\caption{Local DOS  for a chain of  $N=30$ sites, evaluated at sites $i=1,2$ and 15 (panel a, b and c, respectively) as a function of energy and time, following a sudden change in the coupling parameters at $t=30$. For $t<30$ the couplings are uniform across the chain, $t_i=4$, while at $t=30$ they are abruptly changed to the SSH-type configuration with alternating values $(t_1,t_2)=(1,4)$. The on-site energies are set to $\varepsilon_i=4.4$, $\Gamma_i=0.25\Gamma_0$, the arrows denote the energy spacing between the main LDOS peaks.}  
	\label{fig_time1}
\end{figure}
%%%%%%%%%%%%%%%%%%%%%%%%%%%%%%%%%%%%%%%%%%%%%%%%%%%%%%%%%%%%%%%%%%%%%%
%
First, we analyze how the local DOS evolves over time at individual atoms in the chain. Figure~\ref{fig_time1} shows the local DOS dynamics for a chain composed of $N=30$ atoms at sites $i=1, 2$, and in the middle of the chain at site $i=15$. As seen, until the perturbation occurs, for $t<30$, the system remains in a stationary state characterized by $N$ peaks in LDOS function of varying intensity. Note that here $t_i = 4$, and the on-site energies are $\varepsilon_i = 4.4$ (corresponding to charge oscillations with a period of $M = 5$ atoms in the SSH chain). Consequently, the LDOS energy range in the normal chain is $\varepsilon_i \pm 2t_i = (-3.6, +12.4)$, meaning that part of the LDOS lies below the Fermi energy and the sites are therefore partially occupied. At time $t=30$, a sudden change in the coupling parameters between atoms causes a non-equilibrium processes and initiates the system’s evolution toward a new stationary state. 
A topological state forms at the end atoms of the SSH chain at energy $E=\varepsilon_{i}$, as shown in the upper panel. It emerges from the LDOS peaks of the normal chain located near the center of the energy band, while the remaining LDOS peaks evolve into two sidebands of the chain, whose intensity is significantly lower than that of the topological state. The system reaches a stationary state after approximately 20 time units.
The LDOS dynamics observed at the second atom (panel b), as well as at other atoms including the central atom (panel c), reveal the formation of an energy gap in the SSH chain with a width of $2 |t_1-t_2|$ and two energy bands (sidebands), each with a width of $2 \min(t_1,t_2)$. It is worth noting that as a result of the sudden change of the couplings, $t_i$, the LDOS of the normal chain temporarily extends in time into the energy gap region, where it gradually decays with diminishing oscillations over time. 
Furthermore, the formation of LDOS sidebands does not proceed smoothly; rather, it exhibits pronounced oscillatory behavior, which will be discussed in greater detail later.

%
%%%%%%%%%%%%%%%%%%%%%%%%%%%%%%%%%%%%%%%%%%%%%%%%%%%%%%%%%%%%%%%%%%%%%%
\begin{figure}[tb]
	\begin{center}
		\includegraphics[angle=0,width=0.9\columnwidth]{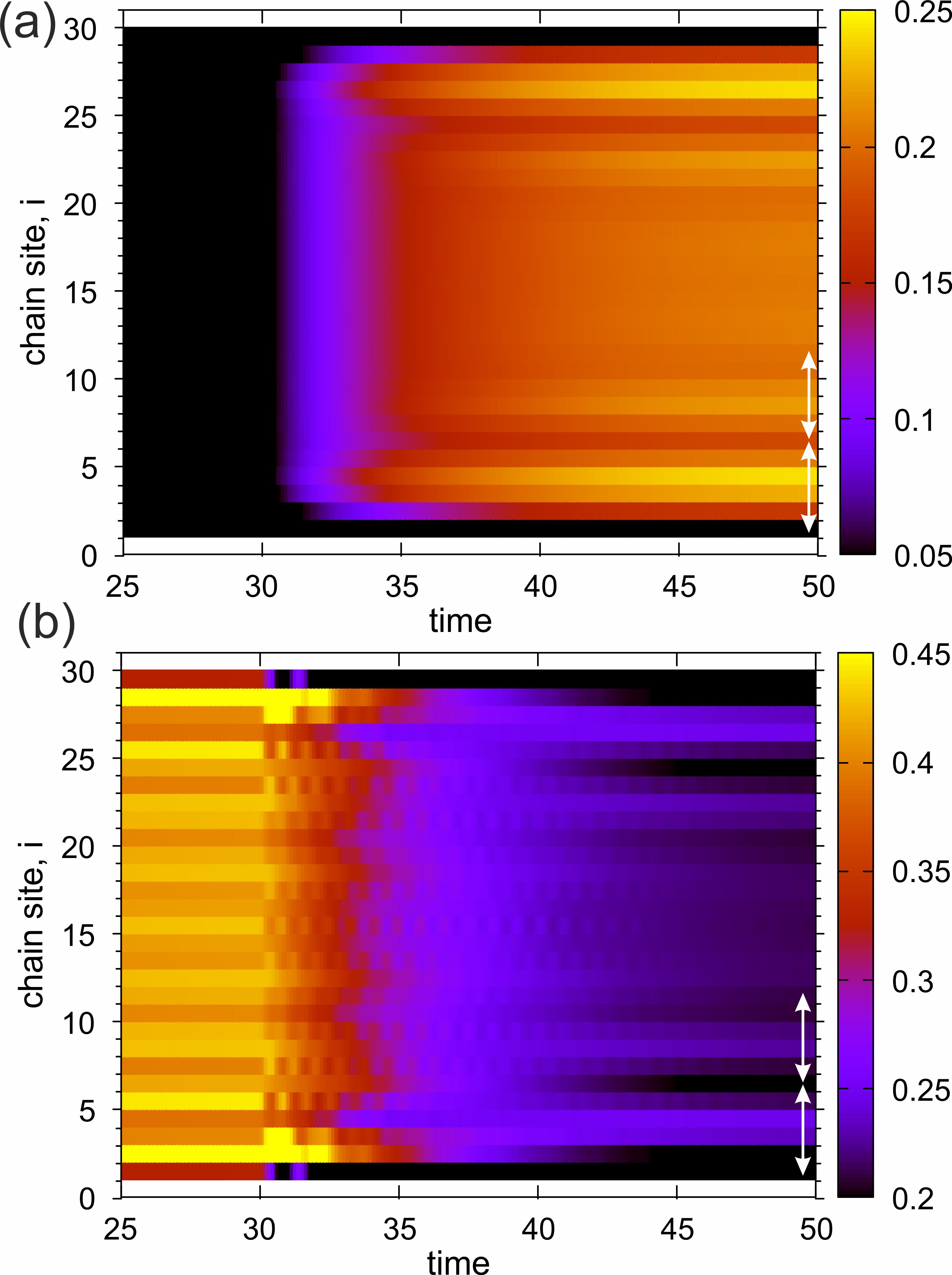}
	\end{center}
	\caption{Charge occupancies, $n_i(t)$ at each chain site $i=1,...,30$  as a function of time for a sudden change in the coupling parameters at time $t=30$. For $t<30$ the couplings parameters along the chain are uniform $t_i=2$ (panel a) and $t_i=4$ (panel b), and for $t=30$, the couplings are changed to those of the SSH1 chain with $(t_1,t_2)=(1,4)$. The on-site energies are set to $\varepsilon_i=4.4$ which corresponds to the oscillation period of $M=5$ atoms (indicated by white arrows), $N=30$, $\Gamma_i=0.25\Gamma_0$.  }
	\label{fig_time2}
\end{figure}
%%%%%%%%%%%%%%%%%%%%%%%%%%%%%%%%%%%%%%%%%%%%%%%%%%%%%%%%%%%%%%%%%%%%%%
%
Knowledge of the time-dependent dynamics of the local DOS at each atomic site allows for the analysis of the occupations. Figure~\ref{fig_time2} presents 3D heat maps of charge variations at all chain sites following a sudden change in the system parameters, which drives the system from a normal state with $t_i=2$ in panel $(a)$ and $t_i=4$ in panel $(b)$ to a topological SSH1 phase with alternating coupling values $(t_1,t_2)=(1,4)$.
In the upper panel we observe that before this change at $t=30$, the chain remains unoccupied  as in this case the LDOS  lies entirely above the Fermi level (small $t_i$ parameter and $\varepsilon_i=4.4$). In the case of stronger couplings (panel b, with $t_i = 4$), for $t < 30$ part of the LDOS band extends below the Fermi level (see Fig.~\ref{fig_time1}), leading to various site occupations that fluctuate around a value of approximately 0.4.
After the coupling parameters change, at longer times, the system exhibits spatial charge oscillations along the chain with a periodicity of five atomic sites (indicated by white arrows in both panels), similar to what is seen in Fig.~\ref{charge02}, panel (b). It is worth noting that these charge distributions for large time are essentially the same in both panels in Fig.~\ref{fig_time2}, and the differences in color of 3D maps  result only from different initial occupations of the chain for $t<30$.
\TK{This indicates that the long-time dynamical behavior of the system is independent of the initial conditions (initial charge occupations).}
Significant differences appear in the early-time dynamics immediately after the sudden parameter change. In panel $(a)$, where the chain was initially unoccupied, the filling of individual atoms progresses smoothly in time. In contrast, in panel (b), where the chain was initially partially filled, we observe distinct and regular charge modulations that develop over time. These oscillations require further detailed investigation.

\begin{figure}[tb]
	\begin{center}
		\includegraphics[angle=0,width=0.9\columnwidth]{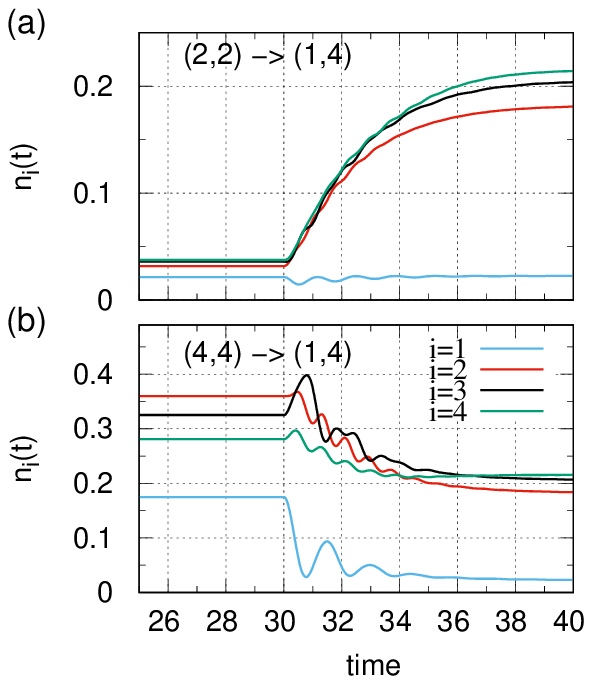}
	\end{center}
	\caption{Time evolution of the charge occupancies, $n_i(t)$  at sites  $i=1,2,3,4$ in a chain of length $N=30$ following a sudden change in the coupling parameters at time $t=30$, as shown in Fig.~\ref{fig_time2}. Initially, the system is a uniform chain with couplings $t_i=2$ (panel a) and $t_i=4$ (panel b), and after the quench, the couplings are changed to those of the SSH1 chain with $(t_1,t_2)=(1,4)$. All parameters are the same as in Fig.~\ref{fig_time2}.}
	\label{fig_time3}
\end{figure}

Figure~\ref{fig_time3} presents a quantitative analysis of the time evolution of charge occupations in a chain following a sudden change in coupling parameters, as discussed in Fig.~\ref{fig_time2}, panels (a) and (b), respectively. We focus on the first four atoms in the chain, as indicated in the legend.
In panel $(a)$ for the coupling change $(2,2)\rightarrow (1,4)$, the charge on the inner atoms increases monotonically up to a value of approximately 0.2 (see also panel $(a)$ in Fig.~\ref{fig_time2}), with only minor time oscillations visible in the curves until a steady state is reached. Notably, the occupation of the first (end) atom behaves differently: after the changing time at $t=30$, it exhibits slight oscillations, but its charge occupation remains very low also at long times. This occurs because the topological state that forms at this site lies entirely above the Fermi energy, while the DOS sidebands have very low intensity (see panel $a$ in Fig.~\ref{fig_time1}) and thus contribute negligibly to the occupation of this state.
%
%
%%%%%%%%%%%%%%%%%%%%%%%%%%%%%%%%%%%%%%%%%%%%%%%%%%%%%%%%%%%%%%%%%%%%%%
%%%%%%%%%%%%%%%%%%%%%%%%%%%%%%%%%%%%%%%%%%%%%%%%%%%%%%%%%%%%%%%%%%%%%%
%
A much more complex dynamical behavior is observed in the charge occupation curves for the case of the coupling change $(4,4)\rightarrow (1,4)$, shown in panel $(b)$. The final values at long times are exactly the same as in panel $(a)$, but the curves here exhibit pronounced temporal oscillations.
We immediately observe that the charge oscillations on the atoms inside the chain ($i=2,3,4$) occur at a higher frequency compared to those on the edge atom ($i=1$, blue curve). These oscillations arise from charge flow along the chain between the neighboring sites, triggered by the perturbation, and are related to Rabi oscillations. These oscillations in a two-level (diatomic) system, occur with a period: $T={2\pi \over \omega}$,   
where $\omega=2 t_{12}$ is energy separation between the system molecular states ($t_{12}$ is the coupling strength between the dimer sites).
In our case, the topological chain features an energy gap bounded by relatively narrow two sidebands, separated by approximately $2\max(t_1,t_2)=8$,  indicated by white arrows in Fig.~\ref{fig_time1}b,c. Consequently, the oscillation period for the inner atoms in the chain is: $T_2={2\pi \over 2\max(t_1,t_2)}\simeq 0.78$. 
At the edge atom the energy separation between relevant levels is nearly twice as small (see the arrows in Fig.~\ref{fig_time1} panel a). As a result, the charge oscillation period at the edge atom is significantly longer than for the inner atoms, and in the considered case is $T_1={2\pi \over \max(t_1,t_2)}\simeq 1.57$. 
\TK{The longer period $T_1$ at the edge atoms arises because the topological edge state is located at the center of the gap, which leads to an energy spacing of approximately $\max(t_1,t_2)$ to the nearest sideband — about half of the spacing relevant for bulk atoms. This is a consequence of the edge state's presence, not a direct measure of the topological invariant.}

It is worth noting that the time-dependent charge occupancy of the third atom in the chain (for $i=3$) exhibits features characteristic of oscillations with both period $T_1$ and period $T_2$. This arises from the fact that the topological state is primarily localized on the last atom of the chain, but it extends over several SSH unit cells \cite{SSHx01} 
\TK{
Consequently, the black curve shows a beating pattern arising from the superposition of two oscillation frequencies, while at site $i=2$ only the shorter-period oscillations are present.
This sublattice-selective penetration of the edge-state signature is a direct consequence of chiral symmetry, which constrains the topological edge state to have weight
exclusively on one sublattice. As a result, the slow oscillation period $T_1$ appears only at odd-numbered sites ($i = 1, 3, 5, \ldots$) with decreasing amplitude, while even-numbered sites
($i = 2, 4, \ldots$) exhibit only the bulk period $T_2$. This alternating spatial pattern distinguishes the topological edge state from a generic boundary-localized state, which would not exhibit such sublattice selectivity. The observation of this pattern in the transient dynamics therefore provides not only evidence for the presence of a mid-gap state, but also for its topological, chiral-symmetry-protected origin.
}

\TK{We emphasize that the two-timescale signature arises from the energy level structure created by the mid-gap edge state, and similar dynamics could in principle result from any boundary-localized state in the gap. However, in clean SSH chains, such edge states are topologically protected: they exist if and only if $\nu = 1$, their energy is pinned to the gap center by chiral symmetry, and they are robust to perturbations preserving this symmetry. The dynamical signature therefore provides a practical method for detecting the presence of these topologically-protected states, and thereby distinguishing the nontrivial from the trivial phase. The signature probes the edge states—which are a manifestation of the bulk topology—rather than the topological invariant directly.}

Figure~\ref{fig_time3}  reveals yet another interesting effect: at the moment the perturbation occurs in the system  the charge occupations of the inner atoms in the chain temporarily increase, while the atom hosting the topological state exhibits a sudden drop in charge. This behavior arises from a reconstruction in the LDOS just after the perturbation.
For the edge atom, the initially broad LDOS associated with the normal chain (see  Fig.~\ref{fig_time1}a) suddenly transforms into a prominent topological state that lies above the Fermi energy. The accompanying side bands that form have significantly lower intensity. As a result, there is a transfer of DOS above the Fermi level, and the charge associated with this state must flow out from this atom at the initial moment.
In contrast, the LDOS on the inner atoms forms strong side bands (see panels b and c in Fig.~\ref{fig_time1}), and since the lower band extends below the Fermi level, this leads to a transient inflow of electrons into the inner atoms. Consequently, the charge occupation curves for the inner atoms increase sharply at $t=30$. 
\TK{The opposite variation of charge on atoms in the middle of the chain, compared to those hosting edge states at the boundaries, is another characteristic feature that distinguishes chains with and without topologically-protected edge states.}
%The opposite variation of charge on atoms in the middle of the chain, compared to those hosting topological states at the edges, is another characteristic feature that distinguishes the chain’s topological phases.

%
%%%%%%%%%%%%%%%%%%%%%%%%%%%%%%%%%%%%%%%%%%%%%%%%%%%%%%%%%%%%%%%%%%%%%%
\begin{figure}[tb]
	\begin{center}
		\includegraphics[angle=0,width=0.95\columnwidth]{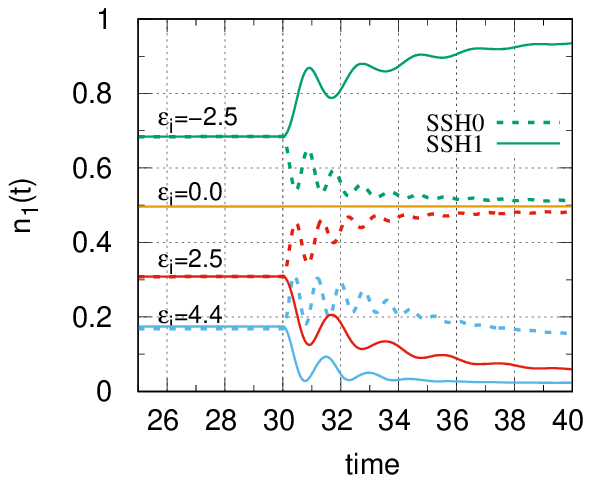}
	\end{center}
	\caption{Time evolution of the charge occupancy at the end atom, $n_1(t)$, following a sudden change in the coupling parameters at time $t = 30$, in a chain of length $N = 30$, for different on-site energies $\varepsilon_i = -2.5$, 0, 2.5, and 4.4 (green, yellow, red, and blue lines, respectively). Solid (dashed) lines correspond to a transition from a normal chain to the SSH1 (SSH0) topological phase, i.e., the couplings change from $t_i = 4$ to $(t_1, t_2) = (1, 4)$ or $(4, 1)$, respectively. All other parameters are the same as in Fig.~\ref{fig_time3}.}
	\label{fig_time4}
\end{figure}
%%%%%%%%%%%%%%%%%%%%%%%%%%%%%%%%%%%%%%%%%%%%%%%%%%%%%%%%%%%%%%%%%%%%%%
%
%The issue of the sudden change in charge occupations occurring at the edge atoms of the chain is also analyzed in fig.
We now turn to investigate whether the results discussed above remain valid for different values of the on-site energies, as well as for the topologically nontrivial phase of the chain in which no topological edge states exist. In Fig.~\ref{fig_time4}, we present results for various positions of the on-site energies  of the chain $\varepsilon_{i}$ (green, yellow, red, and blue lines for the values indicated in the figure), which at time $t=30$ undergoes a transition either to the topologically trivial SSH0 phase (dashed lines) or to the nontrivial SSH1 phase (solid lines). 
The transition to the nontrivial phase for positive $\varepsilon_{i}$ is marked by a sudden drop in the occupation number $n_1(t)$, as also observed in Fig.~\ref{fig_time3}. However, in the case of a transition from the normal state to the topologically trivial phase, we observe a sudden increase in charge (dashed lines for $\varepsilon_{i}>0$). This occurs because in the trivial topological phase, the chain exhibits an energy gap along its entire length, including at the edge atoms. As a result, after the abrupt change in couplings, two intense sidebands emerge, and the lower band drops below the Fermi level, increasing the occupation on those atoms.
Moreover, the frequency of the resulting charge oscillations is twice as high as in the case of a transition to the SSH1 chain (also evident in Fig.~\ref{fig_time3} for inner atoms). This implies that, when the system transitions into the topologically trivial phase, all atoms exhibit oscillations with the same frequency and for positive $\varepsilon_{i}$ are characterized by an initial increase in charge occupancy. In contrast, in the topologically nontrivial phase, the charge dynamics at the edge atom occur at a different frequency and display behavior opposite to that of the inner atoms.

\TK{This effect—different oscillation frequencies at edge versus bulk atoms—provides a clear signature distinguishing the nontrivial from the trivial phase. The higher oscillation frequency observed on all inner atoms indicates the presence of an energy gap. The approximately twofold difference in periods between edge and inner atoms provides direct evidence for a mid-gap localized state at the chain boundary. In the SSH1 phase, this state is the topologically-protected edge state guaranteed by bulk-boundary correspondence ($\nu = 1$). In the trivial SSH0 phase, no such edge state exists, and only the short-period bulk oscillations are observed. The dynamical signature thus detects edge state presence, which in turn reflects the bulk topological invariant.}

It is important to note that in the presence of chiral particle-hole symmetry, $\varepsilon_{i}=E_F=0$, the total charge along the chain remains unchanged and does not respond to a sudden change in coupling parameters (yellow line). In such a case, the local DOS on each atom is symmetric with respect to the Fermi energy, leading to an occupation number of exactly $n_i(t)=0.5$. Furthermore, for negative values of the on-site energy $\varepsilon_{i}<0$, all conclusions concerning the frequency of charge oscillations remain valid, with the only difference being the inversion of oscillation amplitudes on each atom (green curves in Fig.~\ref{fig_time4}). Specifically, if for positive  $\varepsilon_{i}$ the charge on the first atom decreased initially, then for negative $\varepsilon_{i}$ it will increase at the initial moment.
This highlights a universal key feature of topological systems: the analysis of the dynamical evolution of charge occupations provides a clear distinction between topologically trivial and nontrivial phases.

%%%%%%%%%%%%%%%%%%%%%%%%%%%%%%%%%%%%%%%%%%%%%%%%%%%%%%%%%%%%%%%%%%%%
\subsection{\label{sec4e} Transient oscillations from local perturbations}
%%%%%%%%%%%%%%%%%%%%%%%%%%%%%%%%%%%%%%%%%%%%%%%%%%%%%%%%%%%%%%%%%%%%

A sudden change of coupling parameters along the entire chain leads to transient effects and charge oscillations in time. In this section, we consider an atomic chain 
on a surface  where only a single coupling parameter between two neighboring atoms is varied in time. 
The key question is whether such a local change - easier to realize experimentally - can also induce effects similar to those observed when the parameters of the entire chain are modified.

\begin{figure}[tb]
	\begin{center}
		\includegraphics[angle=0,width=0.95\columnwidth]{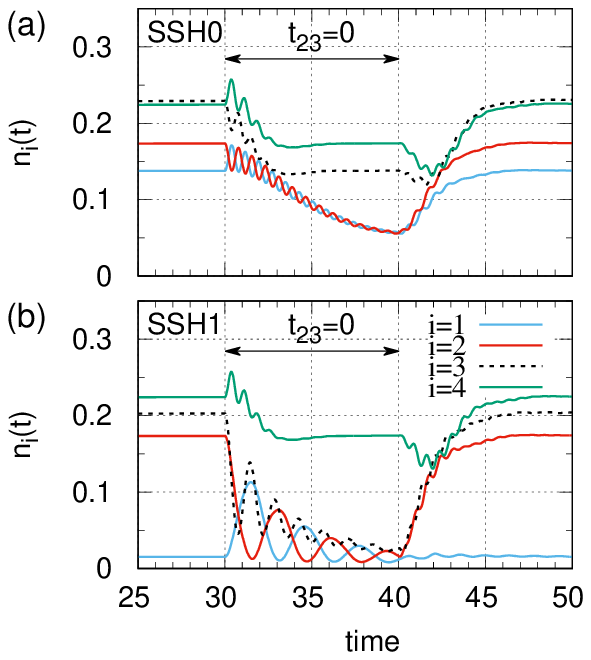}
	\end{center}
	\caption{Time-dependent occupancies $n_i(t)$ at the first four sites ($i=1,2,3,4$) in a chain of length $N=30$, following a sudden change in the coupling parameter $t_{23}$ (between sites 2 and 3), which was set to zero in the time interval from $t = 30$ up to $t = 40$. Initially, the system is in the topologically trivial (panel a) and nontrivial (panel b) phase, with couplings $(t_1,t_2)=(4,1)$ and $(t_1,t_2)=(1,4)$, respectively. The on-site energies are $\varepsilon_i = 4.4$, and all other parameters are the same as in Fig.~\ref{fig_time2}. }
	\label{fig_time5}
\end{figure}

We examine an atomic chain in the topologically trivial phase (panel a in Fig.~\ref{fig_time5}) and in the topologically nontrivial phase (panel b), where electron tunneling between atoms 2 and 3, $t_{23}$, is suddenly blocked within the time interval from $t=30$ to $40$ units. 
We find that the site occupancies respond to this change by developing characteristic transient oscillations in time, as shown in Fig.~\ref{fig_time5} for the first four sites.
In particular, for the SSH0 model, where the system is gapped along the entire chain, the separation between both energy  bands is $2 \max(t_1,t_2)=8$, which yields a transient oscillation period of $T =0.78$. This is clearly visible in the upper panel for the charges on all  atoms of the chain. Notably, the occupancies of atoms 1 and 2 (as well as atoms 3 and 4) oscillate in antiphase, analogous to Rabi oscillations in a diatomic molecule.    

%
%%%%%%%%%%%%%%%%%%%%%%%%%%%%%%%%%%%%%%%%%%%%%%%%%%%%%%%%%%%%%%%%%%%%%%
%%%%%%%%%%%%%%%%%%%%%%%%%%%%%%%%%%%%%%%%%%%%%%%%%%%%%%%%%%%%%%%%%%%%%%
%

The situation changes in the case of a chain in the topologically nontrivial phase, SSH1, where topological states appear at the edge atoms. After switching off the coupling $t_{23}$, the subsystem formed by the first two atoms stands for a dimer, giving rise to Rabi-like charge oscillations with period $T={2\pi \over 2 t_1}=3.14$ (blue and red lines in the lower panel). Atom 4 (green line), both before and after the perturbation, acts as an inner atom of the chain and its LDOS always exhibits two sidebands separated by the energy $2t_2$, which results in transient oscillations with period $T=0.78$. At the same time, atom 3 effectively becomes an edge atom of the rest SSH1 chain. The emergence of a topological state localized on this site leads to charge oscillations with a period determined by half the band gap, i.e., $T={2\pi \over t_2}=1.57$ (black dashed line).
Once the coupling between atoms 2 and 3 is restored at $t=40$, all inner atoms exhibit charge oscillations with a period governed by the full band gap, $T=0.78$. In contrast, atom 1 reconstructs its topological state, so its charge oscillations are associated with half the band gap, corresponding to a period $T=1.57$.
%co jest konsekwencją istnienia fazy nietrywialnej

Thus, in the topologically trivial phase of the chain, transient charge oscillations are solely determined by the existence of the band gap along the chain, and only short-period oscillations are observed. In the nontrivial phase, however, different oscillation periods appear at the edge atoms due to the presence of topological states, together with short-period oscillations on the inner atoms associated with the bulk energy gap. Time-dependent charge dynamics driven by transient effects therefore provide \TK{a sensitive method for distinguishing between the nontrivial and trivial phases by detecting the presence or absence of edge states.}

\TK{
	\textit{Experimental feasibility:} The measurement of charge waves in stationary 1D systems is possible using STM techniques \cite{Jaloch2023}. The dynamical signatures identified in this work should be observable in several experimental platforms. For semiconductor quantum dot arrays with typical parameters $\Gamma_0 \sim 1$~meV and hopping amplitudes $t \sim 0.1$--$1$~meV, the oscillation periods are in the range $T \sim 0.5$--$10$~ps, which are accessible using pump--probe techniques and time-resolved charge sensing. For atomic chains on surfaces with hopping amplitudes $t \sim 0.1$~eV, the relevant timescales are $T \sim 10$--$40$~fs, within reach of ultrafast scanning tunneling microscopy. Cold-atom realizations of SSH chains offer particularly favorable conditions, with oscillation periods in the microsecond-to-millisecond range and direct site-resolved imaging capabilities~\cite{x007}. The key experimental observable — different oscillation frequencies at edge versus bulk sites following a quench — provides a robust, qualitative distinction between topological phases. 
}
%
%%%%%%%%%%%%%%%%%%%%%%%%%%%%%%%%%%%%%%%%%%%%%%%%%%%%%%%%%%%%%%%%%%%%
\section{\label{sec8} Conclusions}
%%%%%%%%%%%%%%%%%%%%%%%%%%%%%%%%%%%%%%%%%%%%%%%%%%%%%%%%%%%%%%%%%%%%
%

In this work, we have explored the formation of charge waves in one-dimensional SSH topological chains.
The appearance of charge oscillations along the chain requires either
(a) explicit breaking of chiral symmetry (e.g., by introducing different on-site energies for alternating atoms), which enforces asymmetry in the LDOS and inevitably leads to even-odd charge oscillations, or
(b) modulation of the LDOS along the chain, regardless of whether chiral symmetry is preserved. In such a case, one of the sidebands must cross the system’s Fermi energy. The resulting oscillation period, which can take arbitrary values, is determined by the average charge filling of the chain.
Our results demonstrate that, despite the presence of an energy gap in the topological phase, charge waves can still develop. The existence of a topological state is not required for their formation, as oscillations are observed in both the trivial and nontrivial phases. Nevertheless, the occurrence of charge-wave modulations with 
\TK{large amplitude modulations at chain ends reflect edge state spectral weight and are characteristic of the nontrivial phase.}
For a chain with two inequivalent atoms within the SSH unit cell, even--odd charge oscillations always emerge. 
\TK{In such systems, a new class of regular oscillations, namely sublattice charge oscillations with longer periods, can also occur simultaneously with the even--odd oscillations.}

We have also investigated dynamical phenomena, focusing on how charge waves develop over time in response to sudden changes in the coupling parameters. Time-resolved analysis of the local DOS reveals that wave formation follows distinct timescales and mechanisms, depending on the initial occupation of the chain. This quench process exposes significant differences depending on whether the system evolves into a trivial or a nontrivial phase. In the trivial phase, all atoms exhibit oscillations with the same frequency and a uniform increase in occupancy.  \TK{In the nontrivial phase, edge atoms display distinct response — oscillations differ in both frequency and phase from bulk atoms, directly reflecting the presence of the mid-gap edge state.} 

Moreover, we show that the transient charge dynamics induced by a sudden change in a single coupling parameter reveal clear distinctions between the trivial and nontrivial phases: in the trivial case, only short-period oscillations of $n_i(t)$ are present, whereas in the nontrivial case, additional oscillation periods emerge at the edge atoms due to the presence of the edge state. 
\TK{Additionally, the sublattice-selective spatial profile of the dynamical signature, where the slow oscillation period appears only on sites belonging to the same sublattice as the edge state,
provides additional evidence for the chiral-symmetry-protected nature of the topological state, distinguishing it from generic boundary effects.}

These results demonstrate that transient dynamics provide \TK{a sensitive means of distinguishing topological phases by detecting the characteristic two-timescale signature of topologically protected edge states.}

%Overall, our findings demonstrate that charge wave phenomena and their dynamics provide a powerful tool for probing topological properties in low-dimensional systems and may serve as a foundation for the experimental detection and control of topological states in nanostructures.

\section*{Acknowledgments}
This work was partially supported by National Science Centre, Poland, under Grant No. 2022/45/B/ST3/01123 (TK), FondeCyT (Chile) under grant number 1250751 (LEFFT), and from the ICTP through the Associates Programme and from the Simons Foundation through grant number 284558FY19 (LEFFT). 
%The data underlying this study are available in Zenodo at \cite{zenodo_data}, https://doi.org/XX.XXXX/zenodo.XXXXXXX. 

\appendix

\section{Details of the Calculations}
\label{app:calc}

\textbf{Green function Method.}
Electron transport through a system described by the stationary Hamiltonian is analyzed here within the framework of Green's functions \cite{Datta,Des}.
The local DOS at each site is obtained from the retarded Green's function $G^r_{ii}(E)$ corresponding to the \mbox{$i$-th} site of the chain, i.e. $LDOS_i(E)= -{1 \over \pi} Im G^ r_{ii}(E)$. The local DOS, in turn, enables the calculation of charge occupations on each site of the chain,  $n_i=\int LDOS_i(E) f(E) dE$, ($f(E)$ stands for the surface Fermi function), making it possible to study charge  waves along the system in the stationary case.
The equation of motion for the Green's function, applied to our $N$-site linear system, leads to the following matrix relation:
$\hat G^r \cdot \hat A = \mathbb{I}$,
where  $\mathbb{I}$ is the identity matrix and  $\hat A$ is a square, tridiagonal  $N \times N$ complex matrix:
%
%%%%%%%%%%%%%%%%%%%%%%%%%%%%%%%%%%%%%
\begin{eqnarray} \label{eq_A}
	A_{i, j}(E) &=& (E
	-\varepsilon_{i})\delta_{i,j} - t_{i,j+1}(\delta_{i,j+1} + \delta_{i+1,j})  \nonumber\\&+& \sum_{\vec{k}}{V^{*}_{i, \vec{k}}V_{j, \vec{k}}}{(E-\varepsilon_{\vec{k}})^{-1}}  .
\end{eqnarray}
% \Sigma_{i,j}(E)
%%%%%%%%%%%%%%%%%%%%%%%%%%%%%%%%%%%%%
The last term in $A_{i,j}(E)$ represents the system's self-energy, which in general takes the form: $\sum_{\vec{k}}{V^{*}_{i, \vec{k}}V_{j, \vec{k}}}{(E-\varepsilon_{\vec{k}})^{-1}}=\Lambda_{ij}(E)-i\Gamma_{ij}(E)/2$, where the real and imaginary parts, $\Lambda(E)$ and $\Gamma(E)$, are connected via the Hilbert transform:
$\Lambda_{ij}(E)= {1 \over 2 \pi} \int_{-\infty}^{\infty} {\Gamma_{ij}(E') \over E-E'} dE'$. 
The self-energy depends exponentially on the atomic distance between sites $i$ and $j$ \cite{Kohler2010, Newns}. As a result, the off-diagonal terms of this function decay rapidly, leading to a model with localized electron states in the electrodes. 
%- valid in the limit of a large product of the electrode Fermi vector and the lattice constant. This assumption holds well for semiconducting or insulating surfaces.
%
The diagonal terms of the self-energy  can be determined from the knowledge of $\Gamma_{i}(E)= 2\pi \sum_{\vec{k}} |V_{i,\vec{k}}|^2   \delta(E-\varepsilon_{\vec{k}})= 2\pi |V_{\vec{k}}|^2 DOS_{i}(E)$, where $DOS_{i}(E)$ denotes the energy-dependent DOS of the surface/electrode coupled with $i$-th site of the chain.  
%Note that knowledge of the surface DOS allows for the computation of both $\Gamma(E)$ and $\Lambda(E)$, and consequently the retarded Green's function, from which the local DOS on each atom of the chain can be determined.
%The local DOS, in turn, enables the calculation of charge occupations on each site of the chain,  $n_i=\int LDOS_i(E) f(E) dE$, ($f(E)$ stands for the surface Fermi function), making it possible to study charge  waves along the system in the stationary case.
%
The calculations simplify under the assumption that the surface DOS is $\vec{k}$-independent, which leads to the wide-band approximation (WBL).
This approximation is valid for a locally flat substrate DOS, free of energy gaps or sharp peaks.
In this case, both components of the self-energy become energy-independent: $\Lambda_{ij}(E) = 0$ and ${\Gamma}_{ij}(E) = \Gamma_i \delta_{ij}$, where $ \Gamma_i$ stands for the effective
chain–surface coupling between $i$-th site of the chain and the electrode. 
Note that for a regular non-topological chain with uniform hopping integrals $t_1=t_2$ and uniform on-site energies $\varepsilon_i$, analytical expressions for the retarded Green's function can be obtained \cite{Kwap2011,Kohler2010,anal2001,Jaloch2023}.
In contrast, for a topological SSH chain, simply form for $G^ r_{ii}(E)$ does not exist, but recursive formulas can be derived that allow for the analytical calculation of the Green's function for arbitrary chain length $N$:
%%%%%%%%%%%%%%%%%%%%%%%%%%%%%%%%%%%%%
\begin{eqnarray} \label{eq_rec}
	G^r_{ii}(E) &=&  {(\alpha \det A_{i-2}-\beta^2_{i-1}\det A_{i-3}) }  \\
	&\times& { {  \det A_{N-i-1}-{\beta^2_{N-i} \over \alpha }\det A_{N-i-2} } \over \det A_{N-1}-{\beta^2_{N} \over \alpha }\det A_{N-2} }   \nonumber .
\end{eqnarray}
% \Sigma_{i,j}(E)
%{ {\texttt{cof} A_{i,i} } \over {\det A_i}  } =
%%%%%%%%%%%%%%%%%%%%%%%%%%%%%%%%%%%%%
Here $A_i \equiv A_{i,i}(E)$ denotes a square $i \times i$ matrix (shown nin Eq.~\ref{eq_A}), with initial conditions  $\det A_0 = 1$ and $\det A_1 = A_{1,1}$.
The recursive relation involves parameters $\alpha = A_{1,1}$ and $\beta_i$, where $\beta_i = t_1$ for even $i$, and $\beta_i = t_2$ for odd $i$, respectively. 
%Therefore, for an SSH chain of the length $N$, it is in principle possible to derive an analytical expression for the Green's function, however, its form is rather complex and lacks transparency.
%, making it less suitable for direct physical interpretation.  

\textbf{Evolution Operator Method.}
The time variations in charge occupation resulting from a temporal perturbation in the interatomic couplings, $t_{i,i+1}(t)$, as well as the corresponding dynamics of the LDOS at atomic sites, are calculated in this paper within the interaction picture using the evolution operator formalism.\cite{33,29,kwap2020} The elements of the evolution operator are obtained by solving the following differential equation:
% \ref{equ:equ1}
%%%%%%%%%%%%%%%%%%%%%%%%%%%%%%%%%%%%%%%%%%%%%%%%%%%%%%%%%%%%%%%%%%%%%%%%%%%%%%%%
\begin{equation}\label{equ:equ1}
	i \frac{\partial }{\partial t} U(t, t_0) = \hat{V}(t) U(t, t_0) \, ,
\end{equation}
%%%%%%%%%%%%%%%%%%%%%%%%%%%%%%%%%%%%%%%%%%%%%%%%%%%%%%%%%%%%%%%%%%%%%%%%%%%%%%%%
% Eq.~\ref{equ:equ1}
where $\hat{V}(t)=U_0(t,t_0) H_{int} U_0^\dagger (t,t_0)$, and $U_0(t,t_0)=\mathcal{T} \exp{\left( i\int_{t_0}^t
	dt' H_0(t') \right)}$. Here $\mathcal{T}$ is the time ordering operator and we assume $\hbar=1$. 
The time-dependent charge occupancy at the $i$-th site of the chain, $n_i(t)$, can be obtained from the corresponding evolution operator matrix elements: \cite{33,Kwap2004,Kwap2011,tsukada1985}:
% \ref{equ:eq2}.
%%%%%%%%%%%%%%%%%%%%%%%%%%%%%%%%%%%%%%%%%%%%%%%%%%%%%%%%%%%%%%%%%%%%%%%%%%%%%%%%
\begin{eqnarray}\label{equ:eq2}
	n_i(t)=\sum_{j=1}^{N} n_j(t_0)|U_{i,j}(t,t_0)|^2
	+\sum_{j,\vec{k}_{j}} n_{\vec{k}_{j}}(t_0)|U_{i, \vec{k}_{j}}(t,t_0)|^2  \, ,\nonumber\\
\end{eqnarray}
%%%%%%%%%%%%%%%%%%%%%%%%%%%%%%%%%%%%%%%%%%%%%%%%%%%%%%%%%%%%%%%%%%%%%%%%%%%%%%%%
where $n_{j}(t_0)$ denotes the initial occupation of the corresponding single-particle state, and since we are not focused on transient effects occurring at initial times, we assume that all chain sites are initially empty, $n_{i}(0)=0$. This implies that the charge occupations within the chain are entirely determined by electrons originating from the substrate/electrodes.
The time evolution of the chain spectral function at each site, i.e., the local DOS, can be obtained from the following relation:
\begin{eqnarray}
	LDOS_i(E,t)=\sum_{\alpha} DOS_{\alpha}(E) | U_{i, \vec{k}\alpha}(t,t_0) |^2 \,, \label{a03}
\end{eqnarray}
where $U_{i, k\alpha}(t,t_0)=\langle i|U(t,t_0)|k\alpha \rangle$, and  $DOS_{\alpha}(E)$ is the lead's/surface DOS.
Using the system Hamiltonian for the same on-site energies in the chain, $\varepsilon_{i}$, and assuming the initial time equal zero, the evolution operator matrix elements required to compute the above physical quantities satisfy the following set of integro-differential equations:
\begin{widetext}
	%%%%%%%%%%%%%%%%%%%%%%%%%%%%%%%%%%%%%%%%%%%%%%%%%%%%%%%%%%%%%%%%%%%%%%%%%%%%%%%%
	\begin{eqnarray}\label{equ:Uik_differential}
		i\frac{\partial U_{i, \vec{k}_{j}}(t)}{\partial t} = \sum_{i'} t_{i,i'}(t) 
		U_{i', \vec{k}_j}(t) + V_{i \vec{k}_j} e^{i(\varepsilon_i-\varepsilon_{\vec{k}_j})t} 
		- i|V_{i\vec{k}_j}(t)|^2\int_0^tdt'  \int d\varepsilon DOS_j(\varepsilon) e^{i(\varepsilon_i-\varepsilon) (t-t')}  U_{i,\vec{k}_j}(t')  
	\end{eqnarray}
	%%%%%%%%%%%%%%%%%%%%%%%%%%%%%%%%%%%%%%%%%%%%%%%%%%%%%%%%%%%%%%%%%%%%%%%%%%%%%%%%
	%
	%where $D_{j}(\varepsilon)$ is the $j$-th lead's spectral density function. 
	This equation represents the most general form of the matrix elements of the evolution operator and must be solved numerically for a given substrate DOS. Equation~\ref{equ:Uik_differential}  constitutes, in fact, a set of thousands of coupled complex integro-differential equations for each $i$ and $\vec{k}$ vector, corresponding to a single time point, $t$. The obtained solution allows for calculating the time-dependent charge distribution along the SSH chain, as well as the evolution of the local DOS at each site.
	Assuming the wide-band approximation, the effective chain–surface coupling can be expressed by the energy-independent function $\Gamma_i$, which allows Eq.~\ref{equ:Uik_differential} to be simplified to the following form:
	%%%%%%%%%%%%%%%%%%%%%%%%%%%%%%%%%%%%%%%%%%%%%%%%%
	\begin{eqnarray}  \label{equ:Uik_differential2}
		{d U_{i,\vec{k}_{j}}(t) \over d t}  &=& -i \sum_{i'} t_{i,i'}(t) U_{i', \vec{k}_{j}}(t)  - i {V}_{i,\vec{k}_{j}} e^{i(\varepsilon_i-\varepsilon_{\vec{k}_{j}})\, t} - {\Gamma_j \over 2}U_{i,\vec{k}_{j}}(t)  \,.
	\end{eqnarray}
	%%%%%%%%%%%%%%%%%%%%%%%%%%%%%%%%%%%%%%%%%%%%%%%%%
	%
\end{widetext}
Note that for a regular chain composed of identical atoms and uniform couplings  along the chain, the set of differential equations (Eq.~\ref{equ:Uik_differential2}) can be reduced via the Laplace transform to a system of algebraic equations and solved exactly. In such a  case, time-dependent analytical expressions for the relevant matrix elements of the evolution operator can be  obtained, enabling the analysis of transient effects for $t \gtrsim 0$ \cite{kwap2020}.
In the case of the SSH chain, performing the Laplace transform is still possible; however, solving the resulting algebraic system analytically leads only to recursive relations (similarly to the stationary Green's function method). This prevents the inverse transform from being carried out and thus makes it impossible to derive closed-form analytical expressions for the matrix elements $ U_{i,\vec{k}_{j}}(t)$.
Moreover, in our setup we do not focus on initial transient effects - instead, the time dependence of the parameters is introduced after the system has stabilized in time, i.e., at larger time values.

\section{SSH chain geometry}
\label{app:geometry}
%
%
%%%%%%%%%%%%%%%%%%%%%%%%%%%%%%%%%%%%%%%%%%%%%%%%%%%%%%%%%%%%%%%%%%%%%%
\begin{figure}[tb]
	\begin{center}
		\includegraphics[angle=0,width=0.75\columnwidth]{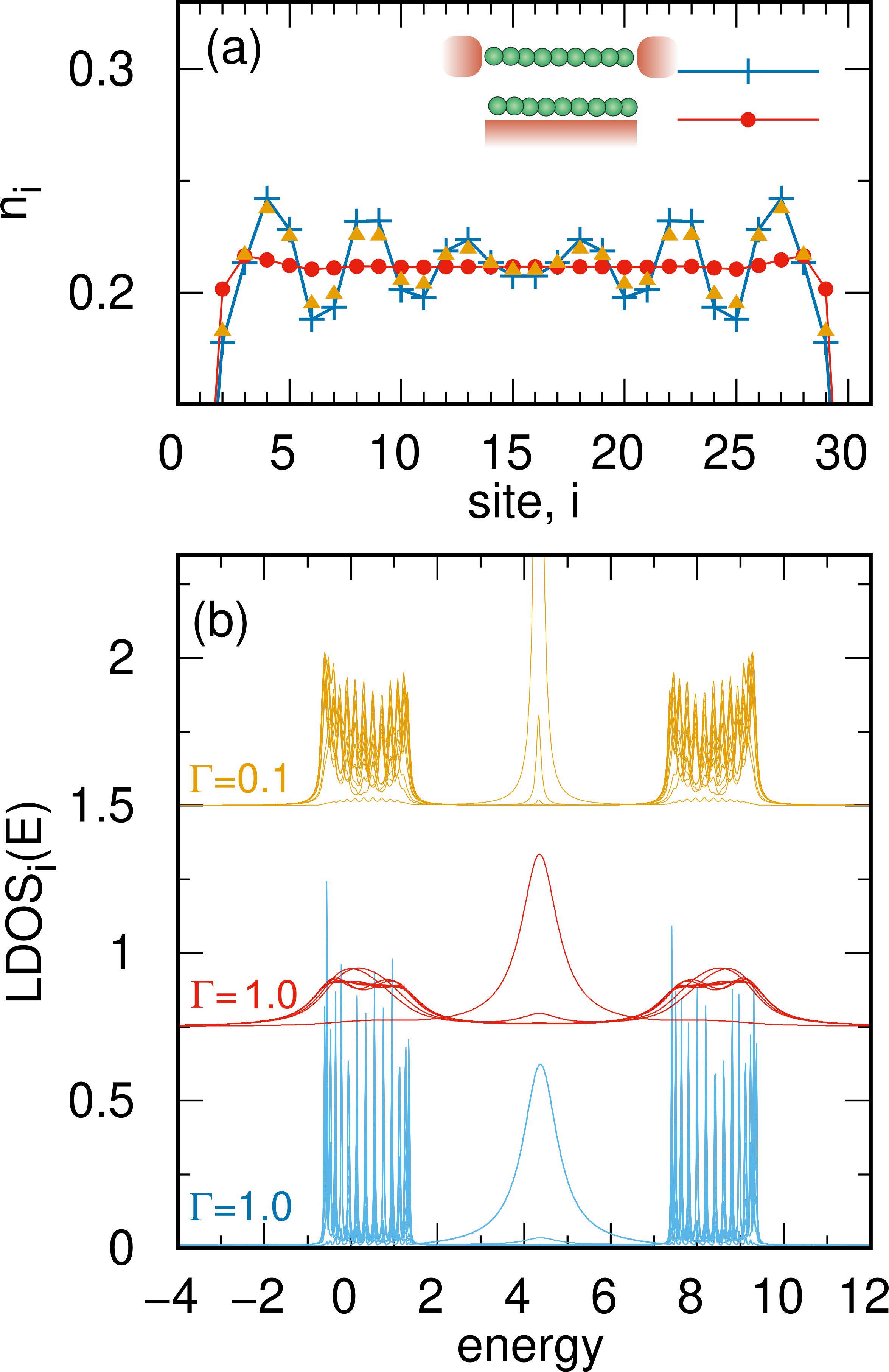}
	\end{center}
	\caption{Charge occupancies (upper panel) along the SSH1 chain of length $N=30$ in the L-R geometry (blue line, $\Gamma_1=\Gamma_N=1$), and for the chain on a surface (red line, $\Gamma_i=1$) for the on-site energies $\varepsilon_i=4.4$. Yellow points in the upper panel represent the occupancies for a chain on the substrate for weak couplings, $\Gamma_i=0.1$. The corresponding local DOS functions at each chain site are depicted in the bottom panel, in the L-R geometry (blue curves), and for a chain strongly or weakly coupled with the surface (red and yellow lines which are shifted by $+0.75$ and $+1.5$ for better visualization); $t_1=1$, $t_2=4$.}
	\label{geometry}
\end{figure}
%%%%%%%%%%%%%%%%%%%%%%%%%%%%%%%%%%%%%%%%%%%%%%%%%%%%%%%%%%%%%%%%%%%%%%
%\end{widetext}
%
The geometry of the system and the nature of the chain–substrate coupling are crucial factors for the observation of charge waves in one-dimensional systems. In the case of linear quantum dot chains, it is common for only the end dots to be coupled to electron reservoirs, which corresponds to the left–right (L-R) geometry. Alternatively, the configuration where all dots are coupled to the electrodes is also possible. For atomic chains assembled on surfaces, each atom typically interacts with the underlying substrate. However, if the substrate is an insulator, applying external electrodes—such as a double-tip scanning tunneling microscope (STM) — to both end atoms effectively realizes the L–R geometry.
The analysis of charge waves for these two geometries is presented in Fig.~\ref{geometry} for the SSH1 chain (of the length $N=30$) exhibiting oscillations with a period of $M=5$ atoms. Charge waves are clearly visible for the L-R geometry (blue line), whereas in the case of a chain fully coupled to the substrate, these oscillations are nearly absent (red line). 
This effect can be explained using the local DOS function, shown in the lower panel. The blue curves correspond to the L-R geometry, and as seen, the LDOS sidebands on each atom exhibit a distinct structure with separated peaks of varying intensity. These peaks are responsible for LDOS modulations in space (along the chain) and for the different occupation at atomic sites. There is also well visible topological state in the middle of the energy gap at $E=\varepsilon_i=4.4$. When each atom is coupled to the substrate, the sideband peaks broaden and overlap, resulting in a relatively smooth local DOS function on each atom. Consequently, the LDOS curves  become very similar at each site (red lines in the lower panel), leading to the suppression of charge waves along the chain.
However, if the coupling of the chain to the substrate is sufficiently weak  the local DOS of the chain closely resembles that of the L-R geometry (yellow curve for $\Gamma=0.1$). Indeed, charge calculations in this case again indicate oscillations with a period of 5 atoms, with almost identical oscillation amplitude (small yellow triangles in panel a).
This suggests that charge waves in topological chains can be observed both in the L-R geometry and in chains interacting with substrates, provided the coupling is weak enough.

\bibliography{bibliography}

\end{document}